# WEAK FOCUSING LOW EMITTANCE STORAGE RING WITH LARGE 6D DYNAMIC APERTURE BASED ON CANTED COSINE THETA MAGNET TECHNOLOGY

A.V. Bogomyagkov, E.B. Levichev, S.V. Sinyatkin

*Budker Institute of Nuclear Physics, Novosibirsk 630090, Russia*

We developed a low emittance electron storage ring with large 6D dynamic aperture. Contrary to the traditional approach using strong focusing magnetic cells with optimized (and large) horizontal phase advance, which yields huge natural chromaticity, we employed a relatively weak focusing lattice with low chromaticity per cell and, consequently, wide on- and off-momentum dynamic aperture. Inevitable for weak focusing emittance growth, we compensated by slicing the lattice into many short, with small bending angle, elementary periodic cells. To reduce the size, we superimposed focusing gradient and chromaticity compensating sextupole components over the dipole field utilizing superconducting magnets based on the Canted Cosine Theta (CCT) winding technology. The result is a model lattice with 50 pm horizontal emittance at 3 GeV beam energy, with ~400-500 m circumference and large 6D aperture.

## I. INTRODUCTION

One of the basic problems of the modern synchrotron light sources with emittances below ~100 pm is a substantial reduction of the 6D dynamic aperture. Meanwhile, a well-proven, effective and simple off-axis horizontal injection needs $A_x \approx 10$ mm; having a good Touschek lifetime in presence of strong intra-beam scattering (IBS) requires large momentum acceptance $A_\delta = (\Delta p/p_0)_{max} \geq \pm 3 \div 5\%$.

Emittance minimization in a synchrotron light source relates to the optimization of horizontal betatron and dispersion functions in the bending magnets, which unavoidably increases beam focusing (or the betatron phase advance over the lattice cell), natural chromaticity and strength of sextupoles required for its compensation. Consequently, the 6D dynamic aperture of such machines shrinks, and in spite of many intensive and shrewd efforts to open it, the problem is still far from resolution [1].

The opposite is also valid: a soft (or relaxed) focusing reduces natural chromaticity and increases dynamic aperture (both transverse and off-momentum). We chose this approach, and to cope with the emittance growth associated with small betatron phase advance, we divided the lattice on the large number of the short and weakly bending elementary periodic cells. This worked, because horizontal emittance depends strongly on the bending angle as $\propto \phi^3$. To keep the ring compact, we superimposed dipole, quadrupole and sextupole fields in a single superconducting magnet of the Canted Cosine Theta (CCT) or Double-Helix design [2]. The CCT design allows combination of several magnetic multipoles in a single magnet with strong field of very high quality.

Similar short cell design of a light source with strong superconducting magnets was recently described in [3]; however, our approach differs both in motivation (low betatron tune advance for wide dynamic aperture) and in technical realization (CCT magnets).

In Section II, we discuss parameters (emittance, chromaticity, quadrupole and sextupole strength, etc.) of a storage ring cell with soft focusing and formulate main approaches for the low emittance cell design. Section III describes a basic lattice cell configuration with 50 pm horizontal emittance at 3 GeV. The cell's off-momentum aperture after chromaticity correction is very large $A_\delta > \pm 40\%$. The cell's transverse aperture is also wide $A_{x,y} \approx \pm 12$ mm for $\beta_{x,y} \approx 1.2$ m at the observation point. In Section IV we construct a 50-pm-emittance storage ring from the basic cells of Section III. Despite the lower symmetry (comparing with the single cell) and following dynamic aperture reduction, it is still large for both efficient injection off-axis injection and good Touschek lifetime. The off-momentum aperture is $A_\delta > \pm 10\%$. The ring design is simple; besides the main CCT combine function magnets, there are only few families of resistive dipoles and quadrupoles to match the straight sections with the cells. We did not use any additional sextupoles or octupoles for dynamic aperture optimization. If added, we believe they will help to increase the dynamic aperture even more. The required cell field distribution has pointed us toward using the CCT magnet technology, which we discuss in Section V.

## II.  LOW EMITTANCE IN A SOFT FOCUSING CELL

In relativistic electron storage ring the horizontal emittance defined by equilibrium between synchrotron radiation damping and quantum excitation can be expressed as

$$\varepsilon_x = \frac{C_q \gamma^2}{J_x} F(\mu_{cx}) \phi_c^3, \qquad (1)$$

where $C_q = 0.383$ pm, $\gamma$ is the beam energy in the rest mass units, $J_x$ is the horizontal damping partition number, $\phi_c$ is the cell bending angle and $F(\mu_{cx})$ is the emittance minimization factor ($\mu_{cx} = 2\pi \nu_{cx}$ is the cell horizontal betatron phase advance and tune).

To establish relations between the beam emittance and ring characteristics, we start with a simple FODO cell showing in Fig.2.1.

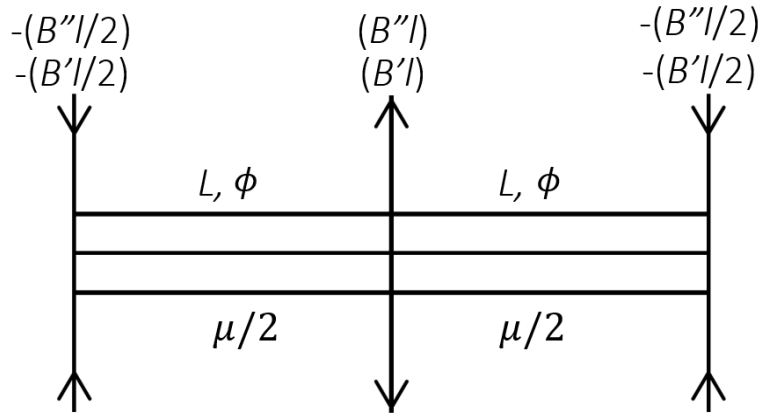

Fig.2.1 FODO cell schematically.

The cell starts and ends with a half-length defocusing quadrupole with overlapped sextupole (both point-like for simplicity). Here we assume $\mu_{cx} = \mu_{cy} = \mu_c$. The bend extends over the entire cell and we neglect its focusing effect. The cell length is $L_c = 2L$ equal double dipole length $L$, the cell bending angle is $\phi_c = 2\phi$. In Attachment, the relevant expressions for both arbitrary and low horizontal phase advance are given. The emittance minimization factor for low phase advance has rather simple form (A1.1)

$$\varepsilon_x \approx \frac{C_q \gamma^2}{J_x} \frac{\phi_c^3}{\mu_c^3}. \tag{2}$$

To keep the constant emittance with the cell tune reduction, one needs to reduce the bending angle proportionally. For the low phase advance, the cell chromaticity is (A5.1)

$$\xi \approx -\frac{\mu_c}{2\pi}. \tag{3}$$

The integrated normalized quadrupole strength is (A4.1)

$$\frac{B'l}{B\rho} \approx \pm \frac{\mu_c}{L}, \tag{4}$$

and normalized integrated sextupole strength needed for chromaticity correction to zero is (A6.1)

$$\frac{B''l}{B\rho} \approx \pm \frac{\mu_c^3}{4L^2 \phi_c}. \tag{5}$$

It seems that both integrated quadrupole and sextupole strengths reduce with the phase advance lowering, however it depends on boundary conditions. In reality, one has different constraints defining the storage ring parameters. Most important are the minimum acceptable emittance (required by users), ring size (coming from the facility cost) and quadrupole and sextupole strength for specified magnet aperture (technical limitation). Keeping the constant emittance $\varepsilon_x$ and the total length of the cells $\Pi_c = L \cdot 2\pi/\phi_c$ (equal to the ring size without drifts, straight and matching sections, etc.) and introducing for brevity the normalized emittance

$$e_x = \frac{\varepsilon_x}{C_q \gamma^2 / J_x}, \tag{6}$$

one may write the following relations

$$e_x \nu_c^3 \approx \left(\frac{\phi_c}{2\pi}\right)^3 = \left(\frac{L_c}{\Pi_c}\right)^3, \tag{7}$$

$$k_1 l = \frac{B'l}{B\rho} \approx \pm \frac{4\pi}{e_x^{1/3} \Pi_c}, \tag{8}$$

and

$$k_2 l = \frac{B''l}{B\rho} \approx \mp \frac{8\pi^2}{e_x \Pi_c^2}. \tag{9}$$

In the low phase advance approximation and for the fixed emittance and circumference of the ring, the integrated quadrupole and sextupole strength *do not depend* on the cell tune, which in turn defines the cell length unambiguously through (7). And vice versa, the cell phase advance *does not depend* on the integrated quadrupole strength. Fig.2.2 shows the integrated quadrupole and sextupole strength as a function of the cell tune for $\varepsilon_x = 100$ pm and $\Pi_c = 200$ m (the cell parameters selection will be discussed in the next section) both for exact (A4, A6) and approximated

(8) and (9) formulas. Note that our simple approximation fits the exact solution rather well up to $v_c \approx 0.4$ that corresponds already the minimum emittance plateau.

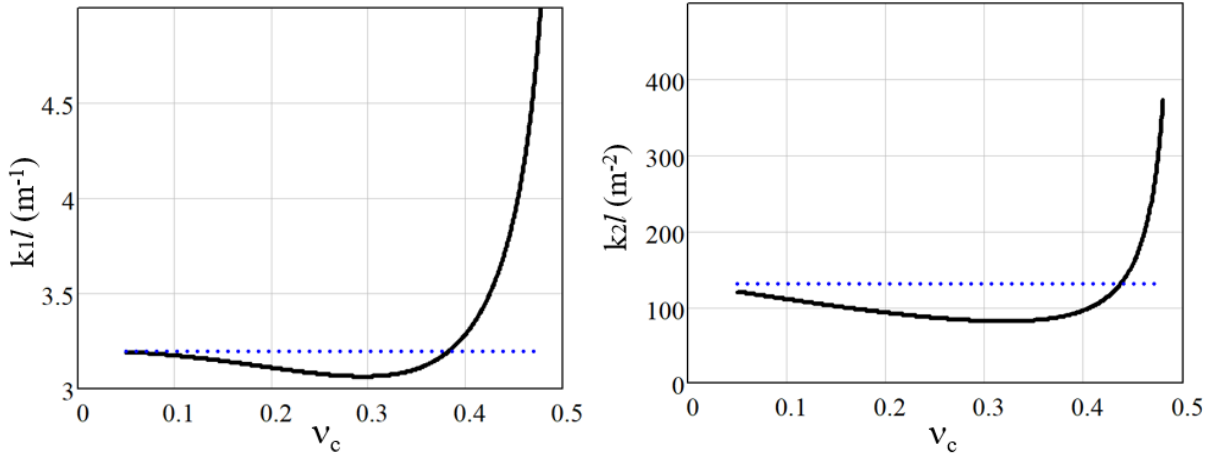

Fig.2.2 Normalized integrated strength of the quadrupole (left) and sextupole magnets. Solid and dashed lines correspond to the exact formulas (A4, A6) and estimations (8) and (9).

But the real technical limitation comes from the field gradient ($B'$ and $B''$) for the specified beam-stay-clear and magnet apertures rather than from the integrated strength. For the cell length $L_c$ and fixed integrated strength, the weaker possible gradient obviously relates to the case when the quadrupole and sextupole fields (we still assume that these components can be superimposed in a single magnet) occupy the half of the cell $l = L_c/2$. Fig.2.3 depicts the gradients as a function of the cell tune for this case. Again, we see that the simple expressions (8) and (9) correspond well to the exact ones (A4, A6) till rather high value of the cell tune.

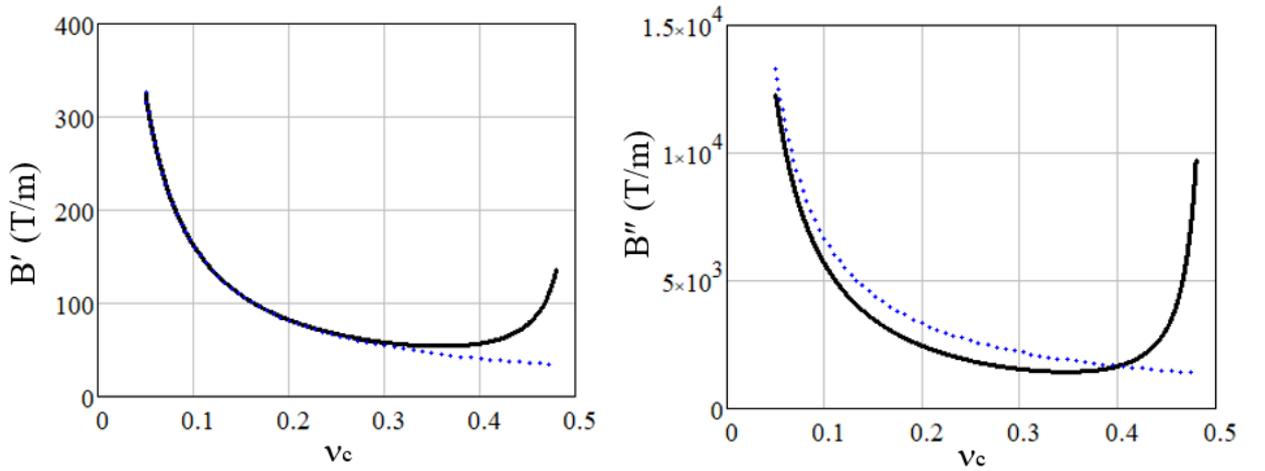

Fig.2.3 Quadrupole (left) and sextupole gradient for $l = L_c/2$ and $B\rho = 10$ Tm ($E = 3$ GeV). As before, the solid and dashed lines correspond to the exact and estimated solutions.

One may notice that for constant emittance and ring size, focusing reduction ($v_c \to 0$) corresponds to *fast increase* of both quadrupole and sextupole gradients (each quarupole and sextupole gradients occupy half the cell length):

$$|k_1| \approx \frac{8\pi}{e_x^{2/3} \Pi_c^2} \frac{1}{v_c} \quad \text{and} \quad |k_2| \approx \frac{16}{e_x^{4/3} \Pi_c^3} \frac{1}{v_c}. \tag{10}$$

Note rather strong dependence of the quadrupole and sextupole gradients on the ring size for specified emittance and horizontal betatron phase advance per cell.

Moreover, vice versa, focusing increase (in terms of the betatron phase advance increase) yields decrease of the gradients with their minimum at the minimum emittance region.

Historically, weak focusing attributes to the azimuthally symmetric accelerator with both low focusing gradient (the field index $0 < |n| < 1$) and betatron tunes $v_x = \sqrt{1-n}, v_y = \sqrt{n}$. In our case, we plan to operate at low tune $v_x < 0.2$ but the focusing gradient will increase with the tune reduction. Therefore, in this paper we mean *the weak* focusing for the lattice cell with low tune (or phase advance) even if the focusing gradient tends to be very high.

Apparently, reduction of the quadrupole and sextupole length below $l = L_c/2$ can only increase the gradient additionally. Another source of the gradients growth in real lattice is optical functions modulation along the cell: in the above estimation, we applied a kick-like magnets placed at the maximum of corresponding betatron and dispersion functions. In addition, if the bending field length is less than the cell length (for instance, there are two drifts of the length $d$ on each side of the bending magnet to accommodate quadrupole or/and sextupole magnets), the emittance for the same horizontal phase advance will grow (in the first approximation order) as

$$\varepsilon_x(d) \approx \varepsilon_x(0)\left(1 + 4\frac{d}{L_c}\right). \tag{10}$$

Now we are ready to formulate the algorithm for the cell parameters evaluation. The energy and emittance are users' constraints. The maximum quadrupole and sextupole gradients are the matter of the cell configuration, magnet technology and beam aperture. Then the cell length and the cell tune are defined unambiguously either from simplified or from exact expressions given in Attachment. Then the question arises, how to find proper tune to get large dynamic aperture and dynamic momentum acceptance (we believe that the lower tune can provide it).

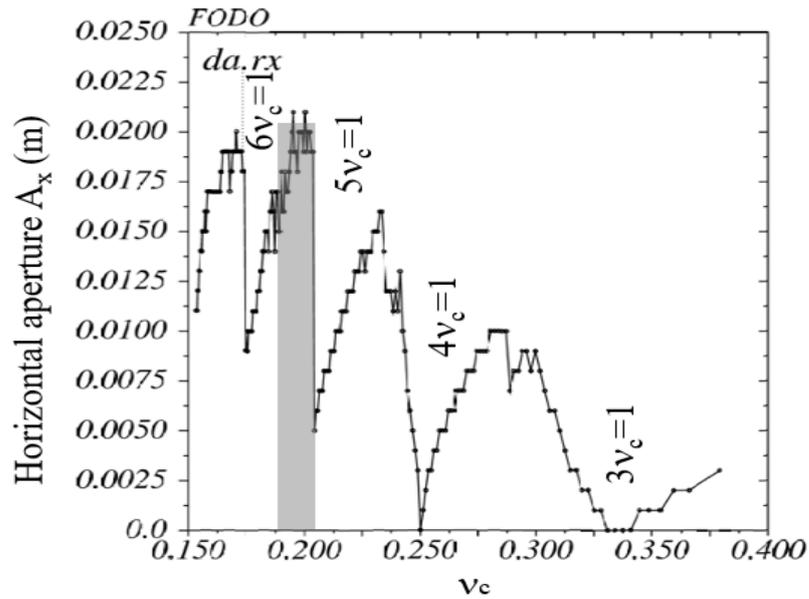

Fig.2.4 Horizontal dynamic aperture of the FODO cell as a function of the tune. Grey bar indicates the cell tune region we apply below.

Detailed analytical and numerical study of the FODO cell nonlinear and chromatic features in [4] confirm our expectation: dynamic aperture indeed opens with the cell tune reduction. Independently, we performed the tracking study for the FODO cell with our parameters (Fig.2.4), with short length and high quadrupole and sextupole gradients.

The results in Fig.2.4 show that there is general tendency of the dynamic aperture increase with the tune decreasing but there are also rather wide dips around the resonant tune values and the cell tune should be defined with considering these dips. A comparison of Fig.2.4 with similar plot of Fig.9 in [4] shows that in our case the dynamic aperture degradation around the resonant lines much more severe due to the stronger sextupoles and higher resonance driving terms.

Low cell advance also allows having large tune bandwidth. According to A7-A10, high order chromaticity

$$\Delta\mu_{x,y}(\delta) = \mu_{x,y}(\delta) - \mu \approx \frac{\mu''_{x,y}}{2}(\delta^2 - 2\delta^3 + \cdots) \approx \mp\frac{\mu}{4}(\delta^2 - 2\delta^3 + \cdots), \quad (11)$$

where the linear chromaticity is corrected by the sextupoles $\xi_{x,y} = \mu'_{x,y} = 0$. According to (11), the nonlinear chromaticity falls down as $\sim\mu$ providing large bandwidth for low cell phase advance.

### III. BASIC CELL DESIGN

A popular class of the fourth generation synchrotron light sources (either new or upgrade of existing machines) has the beam energy of 3 GeV, circumference ~300-400 m (with drifts, straight and matching sections) and natural horizontal emittance ~100-300 pm (see, for instance, a review in [5]). Below we try to design the ring of similar parameters using the approach discussed in the previous section. We aim at the lattice with the cells length of ~200 m and the horizontal emittance of ~100 pm for $J_x = 1$. Adjusting $J_x = 2$ provides additional emittance reduction by factor 2. To reach the main goal of our study – large dynamic aperture/momentum acceptance with chromatic sextupoles only – we choose the horizontal cell tune low enough $\nu_{xc} \approx 0.16 \div 0.17$ according to the grey bar in Fig.2.4. The lower tune increases the quadrupole and sextupole gradients (eqn. (10)) while the higher one reduces the dynamic aperture (Fig.2.4).

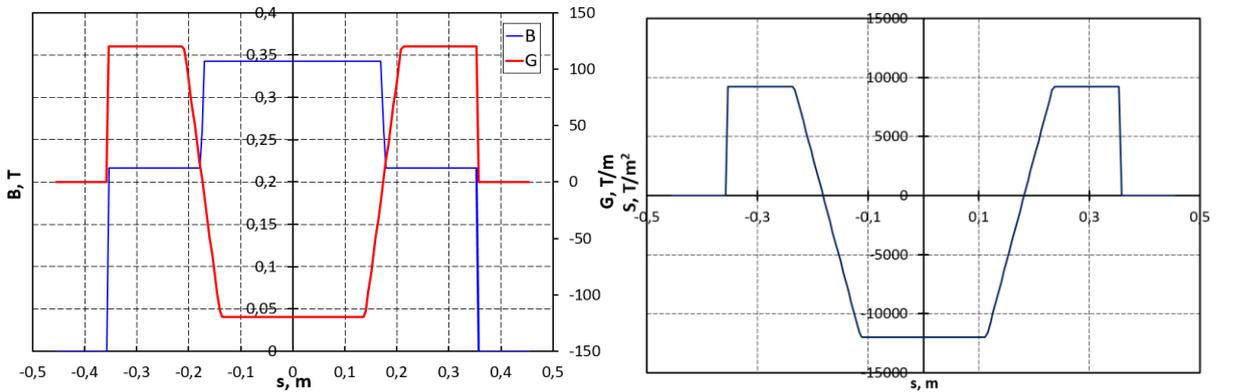

Fig.3.1 Dipole and quadrupole field profile along the cell (left). Sextupole field profile (right).

To develop the cell with the field components overlapping we applied accelerator design code MADX [6]. We fixed the natural emittance $\varepsilon_x = 100$ pm, the total length of the cells as $\Pi_c =$ 200 m, the cell tune as $\nu_{xc} \approx 0.17$, additionally requested zero chromaticity and the bending/gradient field distribution adjusting $J_x = 2$ and ran the MADX optimization module. Fig.3.1 shows the resulting field distribution along the cell.

Fig.3.2 indicates the optical functions for the field pattern from Fig.3.1. Table 3.1 compares the cell parameters, estimated by the expressions given in Attachment compared with those obtained from MADX simulation.

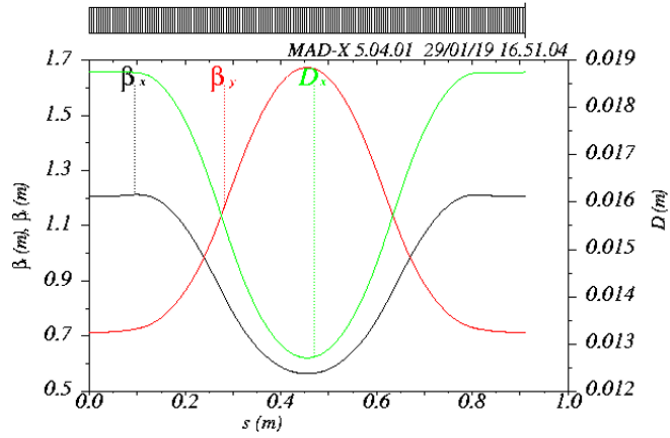

Fig.3.2 Optical functions of the basic lattice cell.

The field and gradient in Fig.3.1 are optimized in such a way to give the partition $J_x = 2$ and reduce the emittance by factor 2. The vertical betatron frequency is tuned slightly down together with the chromaticity compare to the simple estimation in order to achieve larger dynamic aperture in the vertical plane. Other two sources of discrepancies between the estimation and simulation in Table 3.1 relate to the bending pattern (uniform for the estimation and stepwise for the model in Fig.3.1) and to the finite length of the gradient sections (point-like for estimation).

Table 3.1 Basic cell parameters obtained analytically and from the MADX optimization.

|  | Estimation | MADX |
|---|---|---|
| Energy, $E$ (GeV) | 3 | |
| Circumference, $\Pi_c$ (m) | 200 | |
| Emittance, $\varepsilon_x$ (pm) | 100 | 53 |
| Partition $J_x$ | 1 | 1.9 |
| Cell length, $L_c$ (m) | 0.90 | 0.90 |
| Bending angle, $\phi_c$ | 0.021 | 0.02 |
| Tunes, $\nu_{xc}/\nu_{yc}$ | 0.17/0/17 | 0.167/0.149 |
| Natural chromaticity, $\xi_{xc}/\xi_{yc}$ | –0.188/–0.188 | –0.171/–0.159 |
| Momentum compaction, $\alpha \cdot 10^4$ | 4.14 | 3.27 |
| $k_1 l$, (m$^{-1}$) | 3.08/–3.08 | 4.3/–4.08 |
| $B'$, (T/m) | 93.5/–93.5 | 120.9/–115.6 |
| $k_2 l$, (m$^{-2}$) | 184.4/–311 | 326.6/–423.3 |
| $B''$, (T/m$^2$) | 5591.1/–9409.4 | 9225/–11992 |

After some simple optimization of the vertical betatron tune (which is a free parameter), the 4D on-energy dynamic aperture showing in Fig.3.3 was obtained. The aperture observation point corresponds to the initial point of the lattice from Fig.3.1 with the betatron functions $\beta_{x0} = 1.2$ m and $\beta_{y0} = 0.7$ m.

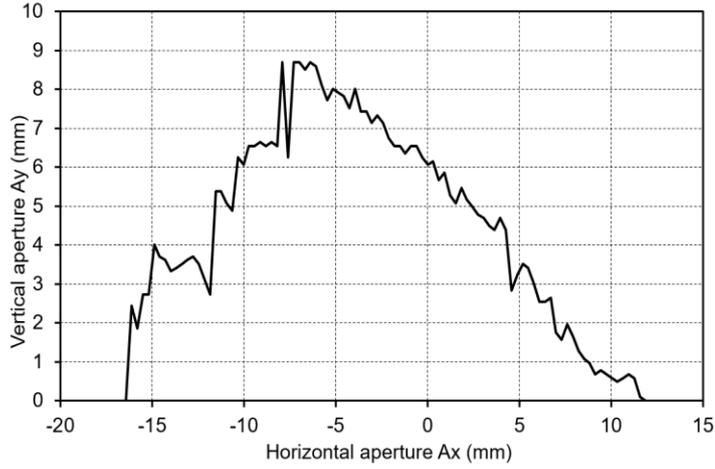

Fig.3.3 Basic FODO cell dynamic aperture. The maximum horizontal aperture reasonably corresponds to Fig.2.4.

The betatron tune bandwidth for the elementary cell is very large. According to the left plot in Fig.3.4 it is in the range of $\Delta p/p_0 = \pm 50\%$, however it is questionable if MADX-ptc_twiss can calculate the bandwidth correctly for such a large momentum variation. Right Fig.3.4 shows the horizontal aperture as a function of momentum deviation. Even for $\Delta p/p_0 = \pm 40\%$ the aperture is still large and again it is not clear can we trust the MADX simulation for so large energy deviation.

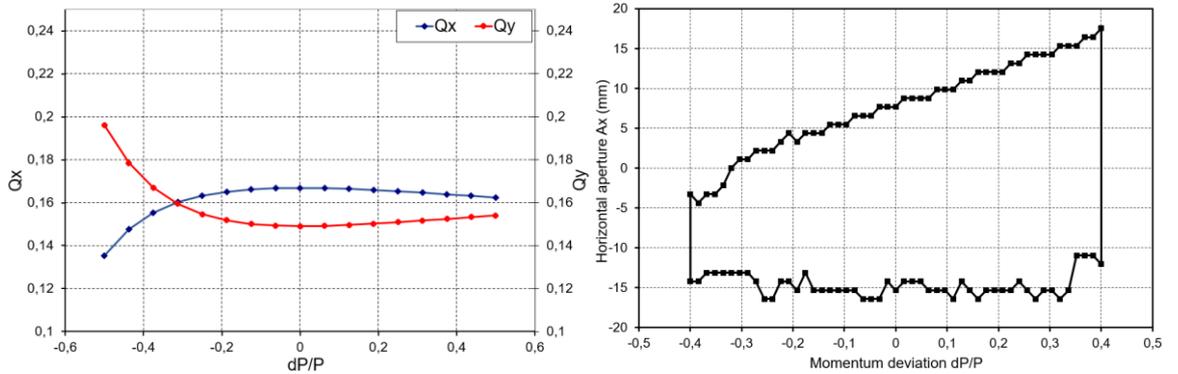

Fig.3.4 Basic cell bandwidth (left) and horizontal dynamic aperture vs momentum deviation (right).

## IV. RING DESIGN

The ring is assembled of 14 super-periods with the length of 28.8 m each. The super-period consists of 22 basic combined function FODO cells with simple matching cells at both ends to remove the dispersion in the straight section. The matching section was optimized to reduce its contribution to the emittance growth but, at the same time, to separate the SR beamline going out from the straight section and the cryostat containing the basic cell magnets. The straight section betatron functions were chosen close to the typical for horizontal injection. Our main goal at this point was to study how the lattice symmetry breaking (as compare to that consisting of the elementary cells only) affects the beam dynamics, dynamical aperture and momentum acceptance.

Fig.4.1 shows the super-period optical functions. Table 4.1 lists main parameters of the ring.

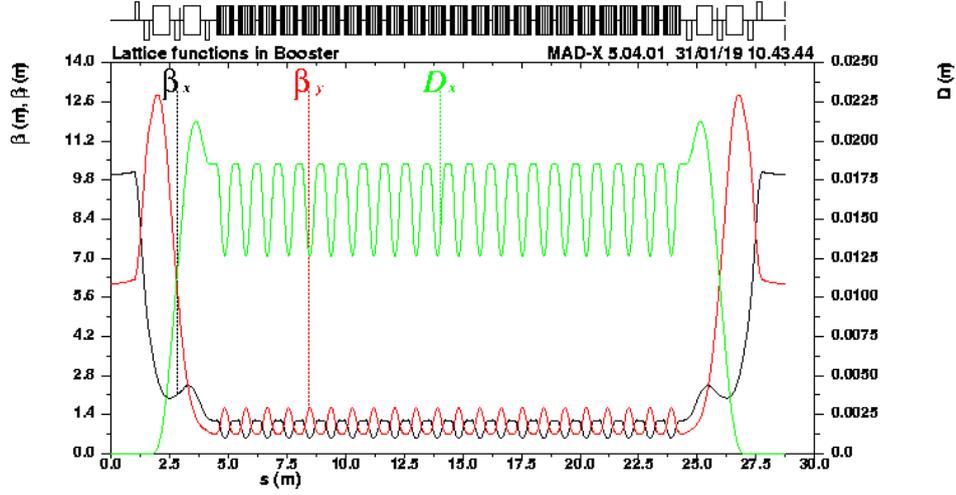

Fig.4.1 Super-period optical functions.

Insertion of the matching and straight sections slightly modified the basic cells optics, increased the natural chromaticity and sextupole strength. We do not use additional multipoles in the matching sections neither to correct the chromaticity no to increase the dynamic aperture. We believe that additional sextupole and octupole families can improve the dynamic aperture.

Table 4.1 Full ring main parameters

| Energy, $E$ (GeV) | 3 |
|---|---|
| Circumference, $\Pi$ (m) | 402.6 |
| Emittance, $\varepsilon_x$ (pm) | 55 |
| Energy spread, $10^4$ | 5.7 |
| Momentum compaction, $\alpha \cdot 10^4$ | 2.2 |
| Energy loss per turn (MeV) | 0.215 |
| Tunes, $\nu_x/\nu_y$ | 58.4/52.9 |
| Natural chromaticity, $\xi_x/\xi_y$ | -59/-74 |
| Partitions, $J_x/J_e$ | 2/1 |
| Damping times, $\tau_x/\tau_e$ (ms) | 21/31 |

The transverse dynamic aperture in the midpoint of the straight section ($\beta_{x0} = 10$ m and $\beta_{y0} = 6$ m) associated with two families of chromatic sextupoles in the basic cells is shown in Fig.4.2. Fig.4.3 shows the full ring betatron tune bandwidth. In spite it is less than that for a single cell in Fig.3.5, it is still larger than ±10%. Fig.4.4 depicts the horizontal off-momentum dynamic aperture, which exceeds ±15%. For this simulation we have been using initial conditions identical for horizontal and vertical planes normalized to the bunch size $N_{\sigma x} = N_{\sigma y}$.

A general conclusion for this section is that despite the transverse dynamic aperture and momentum acceptance reduces compare to the bare basic cell, they are still enough for traditional injection with the stored beam bump in the horizontal plane and for good Touschek lifetime as well.

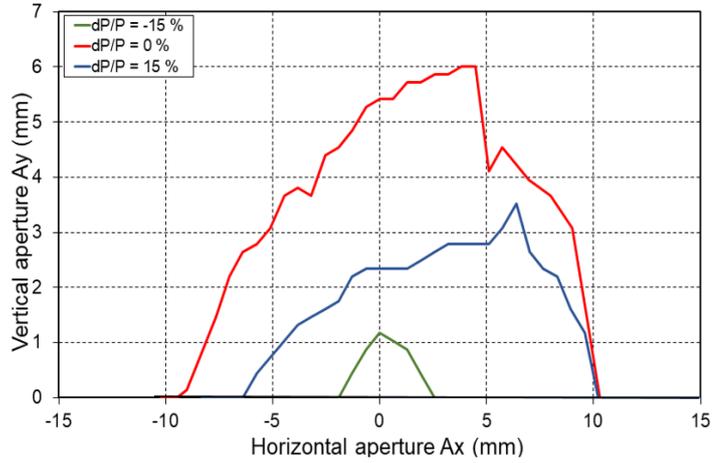

Fig.4.2 Full ring horizontal dynamic aperture for the on- and off-momentum ($\Delta p/p_0 = \pm 15\%$) particles.

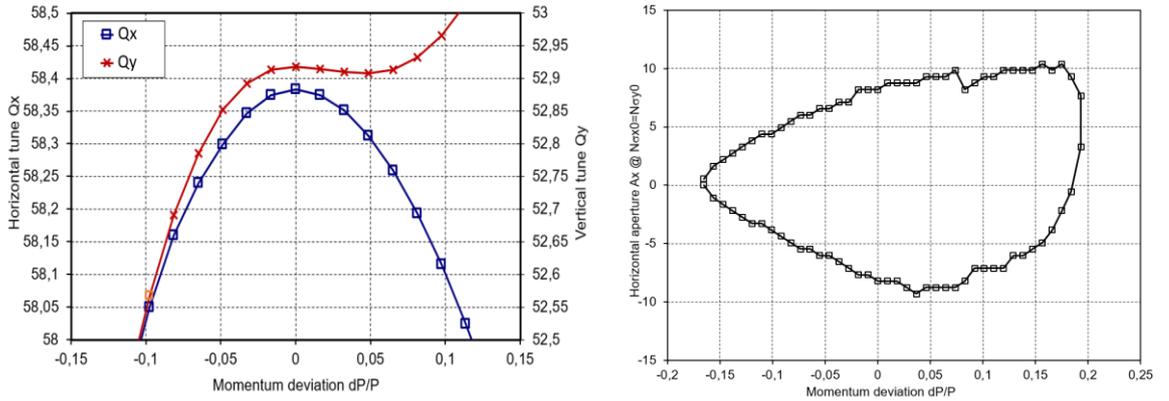

Fig.4.3 Full ring tune bandwidth (left) and horizontal dynamic aperture vs momentum deviation (right).

## V. CANTED COSINE THETA MAGNET FOR BASIC CELL

According to (7), if we reduce the cell betatron phase advance (or the cell tune) and keep the emittance constant, we need either reduce the cell length or increase the cell number (the total ring circumference). The first is difficult for usual separate stepwise magnets while the second yields the tunnel increase and overall facility cost rising. To solve the problem, one may apply a superconducting Double-Helix or Canted-Cosine-Theta (CCT) winding technology [7-9]. The CCT concept is based on even number of conductor layers wound on a cylindrical mandrel (like tilted solenoid) and powered in such a way (Fig.5.1) that their longitudinal field component is cancelled while the transverse is added. Canceled one field component reduces efficiency for the field excitation, which means that more ampere-turns are needed compare to the conventional cosine-theta magnet of the same field. However, for superconducting magnets it is not a big problem while simplicity of design benefits over usual designs.

In the last decade the CCT technology has arouse much interest among accelerator physicists and several superconducting magnets have been designed including NMR coil, heavy ion gantry magnets, high-energy electron-positron collider final focus quadrupoles, high-energy proton-proton collider magnets, etc. [10-12].

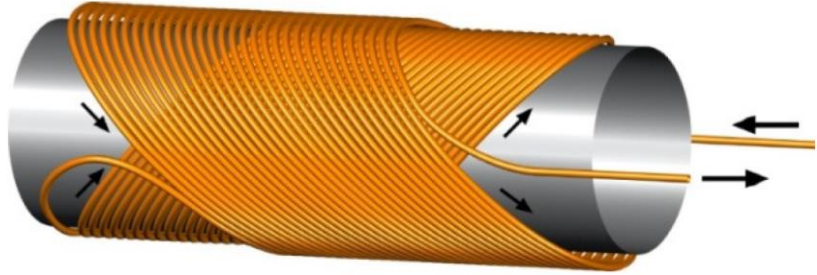

Fig.5.1 Schematic view of two layers of CCT dipole. Proper powering of the conductors cancels the solenoidal field and produces pure vertical field.

Because the conductor distribution forms a perfect sinusoidal current distribution (within machining accuracy), high field homogeneity can be achieved. Fig.5.2 shows schematic CCT winding geometry and notations used below.

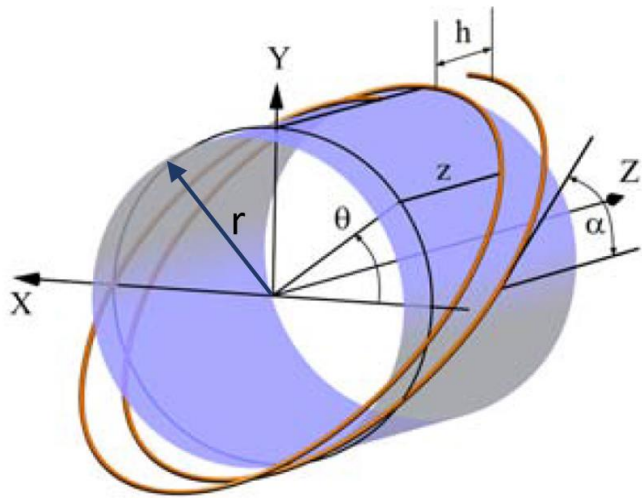

Fig.5.2 Geometry of the CCT winding.

To generate $N$ field harmonics ($n = 1$ corresponds to the dipole field), a three-dimensional curve can be used to define the conductor path expressing as

$$x(\theta) = r \cos \theta, \ y(\theta) = r \sin \theta, \ z(\theta) = \frac{h}{2\pi}\theta + \sum_{n=1}^{N} A_n \sin(n\theta + \varphi_n), \qquad (5.1)$$

where $r$ is the cylinder radius, $h$ is the winding pitch, the phase angle $\varphi_n$ defines normal or skew components and the harmonic amplitude $A_n$ depends on the excitation current.

The conductor placed in precisely machined grooves. A mandrel with grooves is very stable and large Lorenz forces, present in superconducting magnets can be handled very efficiently. Since the magnets are built in a splice-free multilayer system, combined function magnets can be developed within a single winding such as a superimposition of several multipoles and/or twisting or bending. This unique capability is performed without affecting the field homogeneity. Conductor minimum bend radius in the CCT coil is significantly larger than in racetrack or cosine theta magnets. This fact facilitates the use of strains sensitive materials like high temperature superconductors or RRP wires in small dimension magnets. An example of the CCT winding mandrel is shown in Fig.5.2 [14].

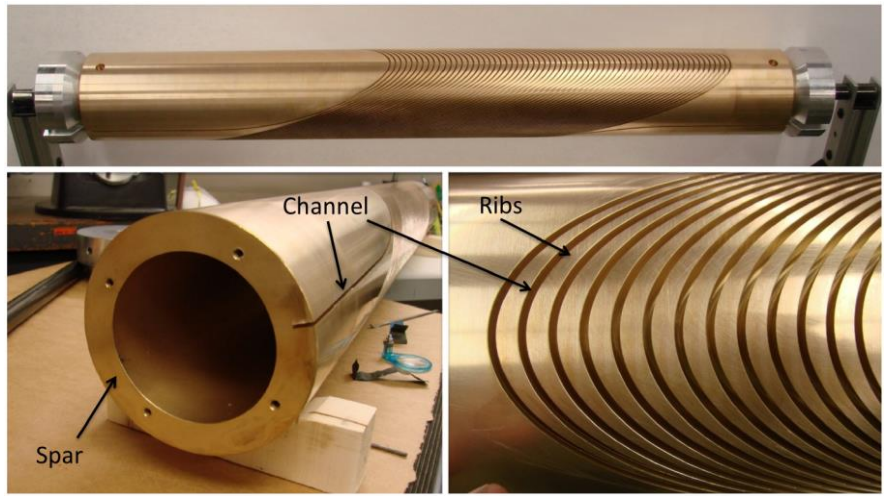

Fig.5.2 A CCT winding mandrel. Rectangular grooves precisely manufactured in a metallic cylinder position Rutherford cable for field quality and wire stability. The spars and ribs provide structural support at the single conductor turn level [14].

In the frame of the CCT technology, the magnets of the basic cell described above are combined in a single two-layer module with the winding modulated along the mandrel as it is shown in Fig.5.3. As here we discuss a possibility of development of CCT combine function magnets, for simplicity the module is straight. However, it is known that bended double helix magnets are also available with minor corrections [15].

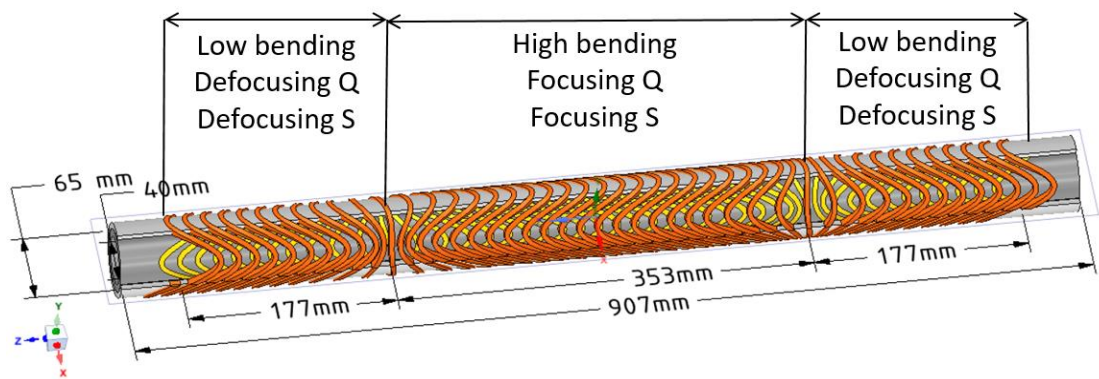

Fig.5.3 Two-layer CCT coil producing the field components according to the above simulation of the basic cell. For illustration purpose, we show here less number of turns with large pitch.

The CCT magnet in Fig.5.3 contains the following segments (see also Fig.3.2 with the field distribution):

- Entry segment to start winding (100 mm).
- Focusing quadrupole and sextupole section (177 mm). Dipole field in this section is low to tune $J_x = 2$.

- Main dipole field segment with the defocusing gradient and sextupole component overlapping (353 mm).
- The symmetrical focusing segment (177 mm) ends the magnet with the grooves finishing part (100 mm).

There are transition areas between the different winding segments where the groove pitch alters the orientation. Fig.5.4 shows the flat projection (evolute) of the currents path with transition area.

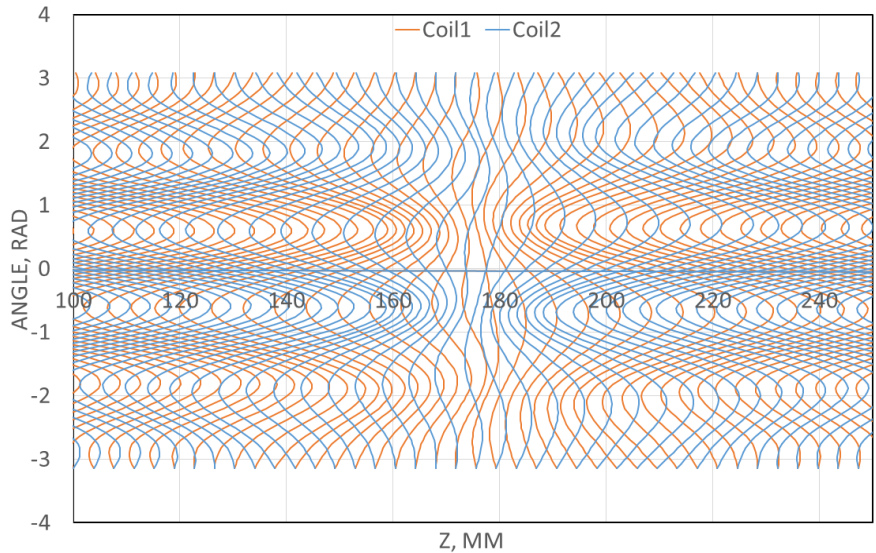

Fig.5.4 Flat projection of the two coils (blue and pink) in the segment transition area.

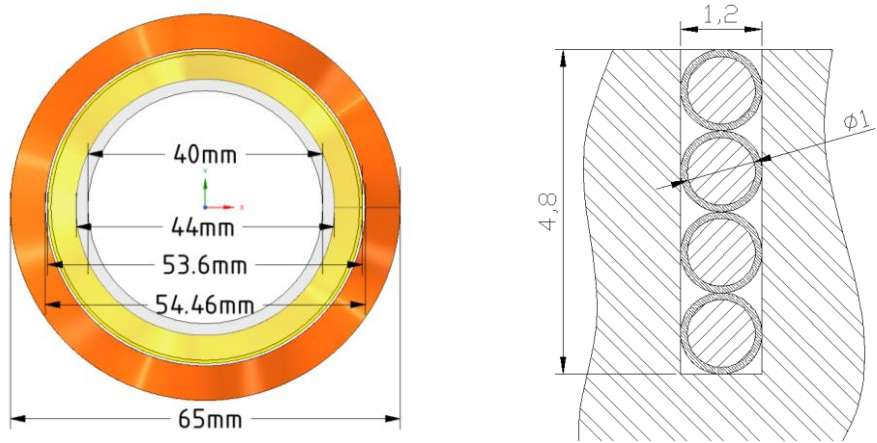

Fig.5.5 A cross section of the CCT magnet to the left. The groove with the superconducting wires (a Rutherford cable can be used as well) to the right.

Fig.5.5 shows the cross section of the CCT magnet mandrels. The outer diameter of the magnet is 65 mm and the inner one is 40 mm where vacuum chamber is located. Two layers (inward and outward) contain either four $Nb_3Sn$ superconducting wires or Rutherford cable in the precisely manufacturing grooves. An excitation current for a single wire is 2.2 kA and the peak magnetic field at the superconductor is 4 T. According to [16-18] at present, a RRP (rod-restack-process) wire allows the critical current density 3000 $A/mm^2$ at 4K and that seems enough for our purpose. The peak field in the superconductor does not exceed 4 T. Table 5.1 lists the main magnet parameters.

Table 5.1 Main parameters of the CCT magnet

| Winding | |
|---|---|
| Current, кА | 2.2 |
| Turns per layer | 4 |
| Layers | 2 |
| Grooves (turns) along the mandrel | 188 |
| Step between grooves, mm | 3.8 |
| Cell length, m | 0.907 |
| **Field parameters in the end segments** | |
| Length, m | 0.177 |
| Dipole field, T | 0.22 |
| Gradient, T/m | 121 |
| Sextupole, $T/m^2$ | 9200 |
| **Field parameters in the central segment** | |
| Length, m | 0.353 |
| Dipole field, T | 0.34 |
| Gradient, T/m | −116 |
| Sextupole, $T/m^2$ | −12000 |

A model of the conductor distribution shown in Fig.5.3 and Fig.5.4 has been developed and the field distribution calculated with the help of Biot-Savart law. The field profile in Fig.5.6 is close to the initial one in Fig.3.2. Further optimization of the current distribution in the mandrels is possible.

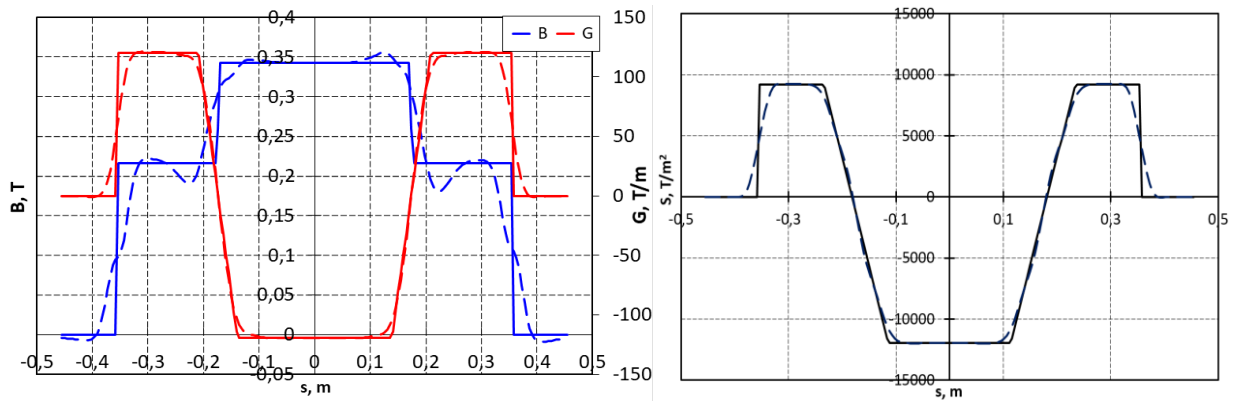

Fig.5.6 The field distribution in the CCT magnet obtained from Biot-Savart law (dash) in comparison with initial model from Fig.3.2 (solid).

## VI. CONCLUSION

Combine function magnets allowing for decrease length or/and enhance performance of particle accelerator. However, conventional magnets mostly have strong leading field component with other(s) week superimposed (strong dipole with low gradient, quadrupole lens with weak bend, etc.). We propose to use the Canted Cosine Theta winding in the superconducting magnets with strong field multipoles superimposed to develop low emittance storage ring with large dynamic aperture. In the example we studied in the paper, we have designed a 3 GeV machine with 50 pm natural horizontal emittance but large on-momentum dynamic aperture $A_x = \pm 10$ mm, which seems enough for conventional off-axis injection. The energy acceptance (and tune bandwidth) is $(\Delta p/p_0)_{max} > \pm 10\%$ providing good Touschek lifetime. Large energy acceptance can be attractive for muon collider lattice also.

**ATTACHMENT**

We present well-known expressions for the FODO cell parameters (see, for instance [19], [20], etc.) for both arbitrary and small cell betatron phase advance ($\mu_x = \mu_y = \mu$). The cell is shown schematically in Fig.2.1; the quadrupoles and sextupoles are point-like while the bending field is extender over the whole cell but its focusing assumed negligible.

| | Exact | | Low phase advance | |
|---|---|---|---|---|
| $e_x = \dfrac{\varepsilon_x}{C_q \gamma^2}$ | $\dfrac{\phi^3}{480} \dfrac{303 + 176\cos\mu + \cos 2\mu}{\sin^3\frac{\mu}{2}\cos\frac{\mu}{2}}$ | A1 | $\left(\dfrac{8}{\mu^3} + \dfrac{1}{2\mu} + \dfrac{7\mu}{120}\right)\phi^3$ | A1.1 |
| $\dfrac{\beta_{max}^{min}}{L}$ | $\pm 2\dfrac{\pm 1 + \sin(\mu/2)}{\sin\mu}$ | A2 | $\dfrac{2}{\mu} \pm 1 + \dfrac{\mu}{3}$ | A2.1 |
| $\dfrac{\eta_{max}^{min}}{L\phi}$ | $\dfrac{2 \pm \sin(\mu/2)}{1 - \cos\mu}$ | A3 | $\dfrac{1}{3} \pm \dfrac{1}{\mu} + \dfrac{4}{\mu^2} \pm \dfrac{\mu}{24}$ | A3.1 |
| $(K_1 L)_F^D$ | $\pm\dfrac{2}{L}\sin\dfrac{\mu}{2}$ | A4 | $\pm\dfrac{\mu}{L}$ | A4.1 |
| $\xi_{x,y}$ | $-\dfrac{1}{\pi}\tan\dfrac{\mu}{2}$ | A5 | $-\dfrac{\mu}{2\pi}$ | A5.1 |
| $L^2 \phi (K_2 L)_F^D$ | $\dfrac{4[\sin(\mu/2)]^3}{\pm 2 + \sin(\mu/2)}$ | A6 | $\pm\dfrac{\mu^3}{4} - \dfrac{\mu^4}{16}$ | A6.1 |
| $\mu_x''$ | $-2\tan\left(\dfrac{\mu}{2}\right)\dfrac{3 + \cos\mu}{7 + \cos\mu}$ | A7 | $-\dfrac{\mu}{2} - \dfrac{\mu^3}{96}$ | A7.1 |
| $\mu_y''$ | $-2\tan\left(\dfrac{\mu}{2}\right)\dfrac{-5 + \cos\mu}{7 + \cos\mu}$ | A8 | $\dfrac{\mu}{2} + \dfrac{13\mu^3}{96}$ | A8.1 |
| $\mu_x'''$ | $-6\mu_x''$ | A9 | $-6\mu_x''$ | A9.1 |
| $\mu_y'''$ | $-6\mu_y''$ | A10 | $-6\mu_y''$ | A10.1 |

According to the notation in Fig.2.1 the cell starts with half of defocusing point-like quadrupole (and the defocusing sextupole locates at the same azimuth). $L, \phi$ and $\mu/2$ relate to the half-cell. In A7-A10 prime denotes derivative with respect to the momentum deviation $\delta = \Delta p/p_0$. In addition, these expressions were derived assuming that the sextupoles compensate linear chromaticity $\xi_{x,y} = \mu_{x,y}' = 0$.